# Partial Reversible Gates(PRG) for Reversible BCD Arithmetic

Himanshu Thapliyal[1], Hamid R. Arabnia[2], Rajnish Bajpai[3], Kamal K. Sharma[4]

[1] NTU, Singapore; [2] The University of Georgia, USA; [3] Synopsys (India) Pvt Ltd, India; [4] G.B. Pant University, India

*Abstract*—IEEE 754r is the ongoing revision to the IEEE 754 floating point standard and a major enhancement to the standard is the addition of decimal format. Furthermore, in the recent years reversible logic has emerged as a promising computing paradigm having its applications in low power CMOS, quantum computing, nanotechnology, and optical computing. The major goal in reversible logic is to minimize the number of reversible gates and garbage outputs. Thus, this paper proposes the novel concept of partial reversible gates that will satisfy the reversibility criteria for specific cases in BCD arithmetic. The partial reversible gate is proposed to minimize the number of reversible gates and garbage outputs, while designing the reversible BCD arithmetic circuits.

## I. INTRODUCTION

Nowadays, the decimal arithmetic is receiving significant attention as the financial, commercial, and internet-based applications cannot tolerate errors generated by conversion between decimal and binary formats. Furthermore, a number of decimal numbers, such as 0.110, cannot be exactly represented in binary, thus, these applications often store data in decimal format and process data using decimal arithmetic software [1]. Since the decimal arithmetic is getting significant attention, specifications for it have recently been added to the draft revision of the IEEE 754 standard for floating-point arithmetic. **IEEE 754r** is an ongoing revision to the IEEE 754 floating point standard [2,3]. It is anticipated that once the IEEE 754r Standard is finally approved, hardware support for decimal floating-point arithmetic on the processors will come into existence for financial, commercial, and Internet-based applications.

Researchers like Landauer have shown that for irreversible logic computations, each bit of information lost, generates $kTln2$ joules of heat energy, where k is Boltzmann's constant and T the absolute temperature at which computation is performed [4]. Bennett showed that $kTln2$ energy dissipation would not occur, if a computation is carried out in a reversible way [5], since the amount of energy dissipated in a system bears a direct relationship to the number of bits erased during computation. Reversible circuits are those circuits that do not lose information and reversible computation in a system can be performed only when the system comprises of reversible gates. These circuits can generate unique output vector from each input vector, and vice versa, that is, there is a one-to-one mapping between input and output vectors. As the Moore's law still holds, the processing power continues to double every 18 months. The current irreversible technologies will dissipate considerable heat and can reduce the life of the circuit. The reversible logic operations do not erase (lose) information and dissipate very less heat. Thus, reversible logic is likely to be in demand in high speed power aware circuits. Reversible circuits are also of high interest in low-power CMOS design, optical computing, nanotechnology and quantum computing. The most prominent application of reversible logic lies in quantum computers [5]. Any unitary operation is reversible and hence quantum networks effecting elementary arithmetic operations such as addition, multiplication and exponentiation cannot be directly deduced from their classical Boolean counterparts (classical logic gates such as AND & OR are clearly irreversible). Thus, quantum arithmetic must be built from reversible logical components.

One of the major constraints in reversible logic is to minimize the number of reversible gate used and garbage output produced. This paper proposes the novel concept of partial reversible gates to minimize the number of reversible gates and garbage outputs, while designing the reversible BCD arithmetic circuits. Thus, an attempt has been tried towards minimizing the number of reversible gates and garbage outputs while designing the optimal reversible BCD arithmetic units.

## II. PARTIAL REVERSIBLE GATES

In this paper, *we propose the novel concept of partial reversible gates which will satisfy the reversibility criteria not in all cases but for specific cases in BCD arithmetic*. We propose this concept to minimize the number of reversible gates and garbage outputs, while designing the complex reversible BCD circuits. Figure 1 shows an example of the proposed concept of partial reversible gate (PRG). Table I shows the truth table of the proposed PRG gate shown in Fig 1. It is to be noted from Table-I that the proposed partial reversible gate (PRG) is reversible only in 'Sect-1', and from 'Sect-2' it loses its feature of reversibility. On carefully examining 'Sect-1' in Table I, we can easily find out that it is the truth table of BCD to excess 3 code converter. The 'Sect-2' of Table-I will never occur in Truth Table of BCD to excess 3 code converter, as BCD number goes only till 1001. Thus, we can use the PRG gate shown in Fig.1 to design reversible BCD to excess 3 code converters since the non-reversible case occurring in 'Sect-2' starting from 1010 in Table-I will never be encountered. The proposed PRG in Fig. 1 is reversible for BCD to excess 3 code converter as there is one to one mapping between input and output vectors which is the primary requirement for logical reversibility.

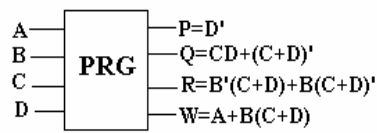

Figure 1. Partial Reversible GateAdvantages of Partial Reversible Gate

## A. Advantages of Partial Reversible Gates

The advantage of the proposed concept of partial reversible gates is that we are able to realize a complete BCD to excess-3 code converter with only one reversible gate and no garbage output. Thus the proposed concept of partial reversible gates will be a boon for designing reversible circuits with minimal number of reversible gates and garbage outputs. *Similarly, other Boolean functions can be examined and partial reversible gates can be proposed for them.* The partial reversible gate concept will be especially beneficial in BCD arithmetic as BCD number only goes from 0 to 9. A case study and comparison of implementing BCD to excess-3 code converter with partial reversible gate and existing conventional method is discussed below.

TABLE I. TRUTH TABLE OF PRG GATE

| A | B | C | D | P | Q | R | S |
|---|---|---|---|---|---|---|---|
| **SECT-1** | | | | | | | |
| 0 | 0 | 0 | 0 | 0 | 0 | 1 | 1 |
| 0 | 0 | 0 | 1 | 0 | 1 | 0 | 0 |
| 0 | 0 | 1 | 0 | 0 | 1 | 0 | 1 |
| 0 | 0 | 1 | 1 | 0 | 1 | 1 | 0 |
| 0 | 1 | 0 | 0 | 0 | 1 | 1 | 1 |
| 0 | 1 | 0 | 1 | 1 | 0 | 0 | 0 |
| 0 | 1 | 1 | 0 | 1 | 0 | 0 | 1 |
| 0 | 1 | 1 | 1 | 1 | 0 | 1 | 0 |
| 1 | 0 | 0 | 0 | 1 | 0 | 1 | 1 |
| 1 | 0 | 0 | 1 | 1 | 1 | 0 | 0 |
| **SECT-2** | | | | | | | |
| 1 | 0 | 1 | 0 | 1 | 0 | 1 | 1 |
| 1 | 0 | 1 | 1 | 0 | 0 | 1 | 1 |
| 1 | 1 | 0 | 0 | 1 | 1 | 1 | 1 |
| 1 | 1 | 0 | 1 | 0 | 0 | 0 | 1 |
| 1 | 1 | 1 | 0 | 1 | 0 | 0 | 1 |
| 1 | 1 | 1 | 1 | 0 | 1 | 0 | 1 |

## B. Case Study

It can be revealed that excess-3 equivalent code can be obtained from the BCD code by the addition of 0011. This addition can be easily obtained through a 4-bit reversible parallel adder. The design of BCD-to-Excess-3 code converter designed using efficient and optimized reversible parallel adder designed from TSG gate [7] is shown in Figure 2. Table II shows the result, that compares the reversible implementation of the BCD-to-Excess-3 code converter using TSG gate, with the proposed concept of partial reversible gate (PRG). It can be observed from Table 2 that implementing the BCD-to-Excess-3 code converter with the proposed concept of PRG has an improvement ratio of 400% and 900% in terms of number of reversible gates and garbage outputs respectively, compared to its implementation with conventional method. Considering the delay of one gate as one unit, there is also a reduction of 400% in unit delay.

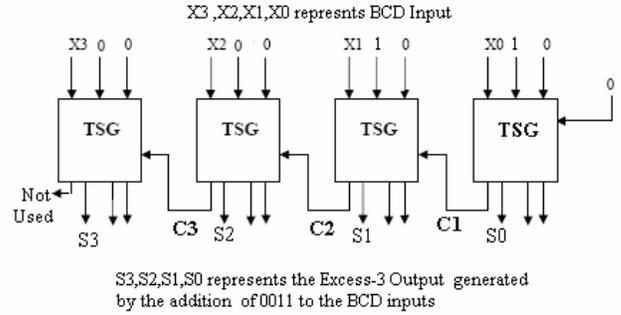

Figure 2. Reversible BCD to Excess 3 Code Converter

TABLE II. A COMPARISON OF IMPLEMENTATION OF BCD-TO-EXCESS-3 CODE CONVERTER

| | Number of Reversible Gates | Number of Garbage Outputs | Unit Delay |
|---|---|---|---|
| Full adder Using TSG | 4 | 9 | 4 |
| Proposed PRG Gate | 1 | 0 | 1 |
| Improvement Ratio | 400% | 900% | 400% |

## III. CONCLUSIONS

The focus of this paper is the design of reversible BCD arithmetic units with minimal gates and garbage outputs considering the growing importance of IEEE 754r (the ongoing revision considering decimal arithmetic). Thus, this paper proposes the novel concept of partial reversible gates which will satisfy the reversibility criteria not in all cases but for specific cases. The partial reversible gates are of great help in minimizing the number of reversible gates and garbage outputs.